\documentclass[
 amssymb, amsmath,%
 aip,cha,%
]{revtex4-1}
\usepackage{graphicx}
\usepackage{bm}%
\usepackage[colorlinks=true,linkcolor=blue]{hyperref}%
\expandafter\ifx\csname package@font\endcsname\relax\else
 \expandafter\expandafter
 \expandafter\usepackage
 \expandafter\expandafter
 \expandafter{\csname package@font\endcsname}%
\fi
\newcommand{\be}{\begin{equation}}
\newcommand{\ee}{\end{equation}}
\newcommand{\ba}{\begin{eqnarray}}
\newcommand{\ea}{\end{eqnarray}}

\begin{document}

\title{Radius-Voltage Relation of Graphene Bubbles Controlled by Gate Voltage}%

\author{Weihua Mu}
\email{muwh@itp.ac.cn, whmu@mit.edu}
\affiliation{State Key Laboratory of Theoretical Physics, Institute of Theoretical Physics, The Chinese Academy of
Sciences, P. O. Box 2735 Beijing 100190, China}
\affiliation{Department of Chemistry, Massachusettes Institute of Technology, Cambridge, Massachusetts 02139, United States}
\affiliation{Kavli Institute for Theoretical Physics China, The Chinese Academy of Sciences, P. O. Box 2735 Beijing 100190, China}
\author{Gang Zhang}
\affiliation{Institute of High Performance Computing, Singapore 138632}
\email{zhangg@ihpc.a-star.edu.sg}

\author{Zhong-can Ou-Yang}
\affiliation{State Key Laboratory of Theoretical Physics, Institute of Theoretical Physics, The Chinese Academy of
Sciences, P. O. Box 2735 Beijing 100190, China}
\affiliation{Kavli Institute for Theoretical Physics China, The Chinese Academy of Sciences, P. O. Box 2735 Beijing 100190, China}
\affiliation{Center for Advanced Study, Tsinghua University, Beijing 100084, China}
\email{oy@itp.ac.cn}

\begin{abstract}
Graphene on the substrate can form bubbles rising above the sheet. In some cases, the bubble is a perfect spherical surface crown, and its radius can be adjusted by external electric field. In this manuscript, we theoretically investigate the voltage dependence of the spherical bubble's radius. The calculated results are in good agreement with recent experiments on the graphene bubble controlled
by applied gate voltage [Appl. Phys. Lett. {\bf 99}, 093103~(2011)].
\end{abstract}
\pacs{61.48.-c, 61.48.Gh, 62.25.-g}
\maketitle
Graphene, the two-dimensional honeycomb structure consisting of a single atomic layer of carbon, is the subject of intensive studies due to its distinctive electronic mobility~\cite{geim07,geim09}, high thermal conductivity~\cite{balandin08,wang11} and high mechanical strength~\cite{frankand07,cranford11,mu09}. With its extraordinary properties, graphene has many electronic~\cite{lin10,mu11}, mechanical~\cite{bunch07}, photonic~\cite{li09,xia09}, thermal~\cite{yang09,hu09,zhang11}, chemical~\cite{ling10}, and acoustic applications~\cite{tian11}. Moreover, graphene is an extremely interesting candidate for micro- and nanoelectromechanical systems (MEMS/NEMS). MEMS/NEMS are devices integrating electrical and mechanical functionality on the same devices. In the design of NEMS devices, the coupling between electrical and mechanical performance is very important. One recent experiment~\cite{georgiou11} found that spherical graphene bubbles can be stable in large size on top of silicon oxide substrate with the bubble radius of graphene spherical bubbles can be controlled by the gate voltage. This may open up graphene's application as optical lenses with adjustable focal length. However, there still lacks an analytical expression of the gate voltage dependence of the bubble radius in the literatures. In the present work, we will fulfill this task.

\begin{figure}
\includegraphics[scale=0.6]{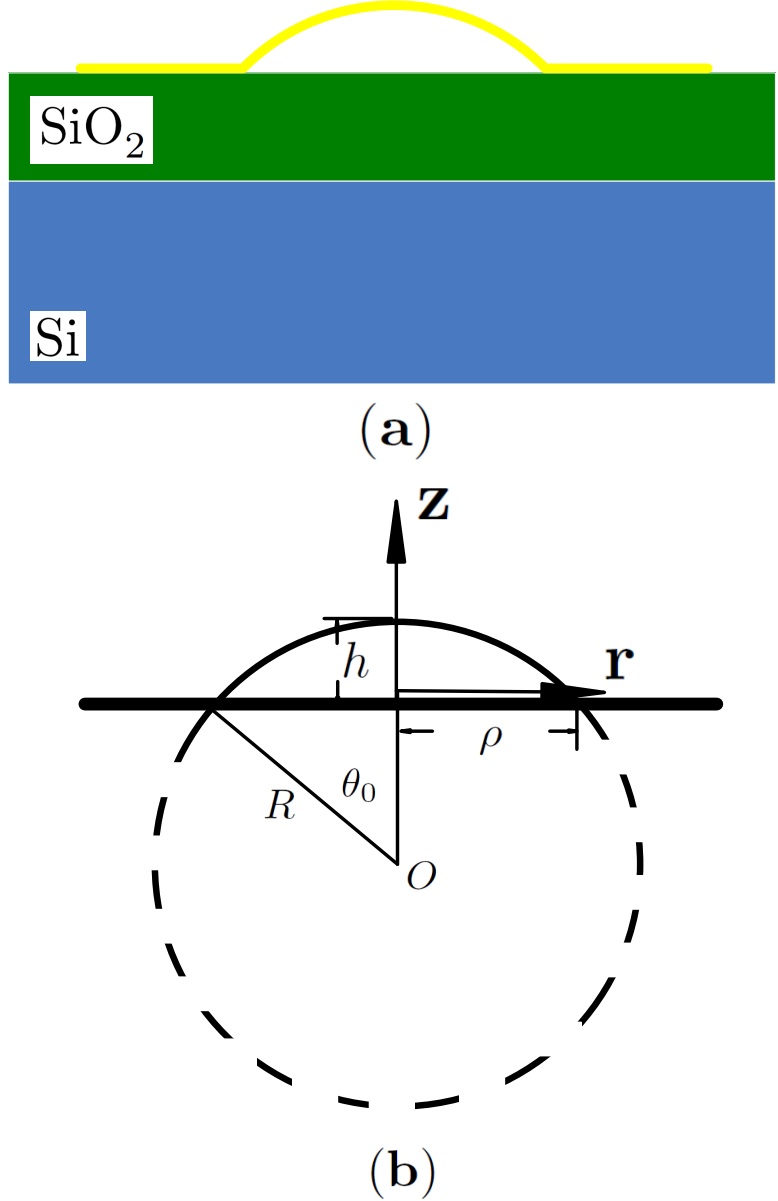}
\caption{\label{fig:fig1}
Schematic illustration of a spherical bubble risen in monolayer graphene. The graphene is on top of the silicon oxide substrate.}
\end{figure}

Among the efforts on the studies of the bubbles rising in graphene sheet~\cite{bunch08, levy10,zong10,koenig11,zabel12,yue2012}, Yue {\it et al.}~\cite{yue2012} proposed a method to theoretically reproduce the shape of a stable graphene bubble observed by Georgiou \emph{et al.}. The equilibrium shape of the bubble is determined by finding the minimum of the free energy containing elastic, adhesion and pressure terms, $F=E_{\mathrm{elas}}+E_{\mathrm{adh}}-\left(p-p_0\right)V$, where $p$ is the pressure of air molecules "sealed" inside the bubble, and $p_0$ is the pressure outside the bubble. We extend this model to describe electromechanical coupling by including the capacitive energy for electrostatic interaction. The main idea of the present work is to introduce a gate voltage dependent correction for adhesion energy term, $c_g V_g^2/2$, with $c_g$ being the Metal-Oxide-Semiconductor (MOS) capacitance per unit area for the graphene/SiO$_\mathrm{2}$/Si structure in Georgiou \emph{et al.}'s experiment, as shown in Fig.\ref{fig:fig1}(a). The expansion and shrinkage of a bubble change the area of the substrate covered by the graphene sheet, thus changing the stored electrostatic energy (capacitive energy) in this MOS system. On the other side, a spherical graphene bubble will expand (shrink) to one with larger (smaller) $h$ and $\rho$, (see Fig.\ref{fig:fig1}(b)) in response to the decreasing (increasing) of gate voltage.

In present work, graphene is modeled as a flexible membrane with only one layer of carbon atoms. The elastic energy can be described by~\cite{yakobson96}
\begin{widetext}
\be\label{elasticenergy}
E_{\mathrm{elas}}=\frac{1}{2}\int\,\left\{ D\left[\left(\kappa_{x}+\kappa_{y}\right)^{2}-2(1-\nu)
\left(\kappa_{x}\kappa_{y}-\kappa_{xy}^{2}\right)\right]+\frac{E_{\mathrm{2D}}}{1-\nu^{2}}\left[\left(\varepsilon_{x}+\varepsilon_{y}\right)^{2}
-2(1-\nu)\left(\varepsilon_{x}\varepsilon_{y}-\varepsilon_{xy}^{2}\right)\right]\right\} \mathrm{d}^2s.
\ee
\end{widetext}
Here, the first term on the right hand side of Eq. (\ref{elasticenergy}) is the strain energy, and the second term is the curvature energy. The $\kappa_{x},\kappa_{y}$ and $\kappa_{xy}$ are local curvatures, $E_{\mathrm{2D}}=Yt=59\,\mathrm{eV/atom}=360\,\mathrm{J\cdot m^{-2}}$, $D=Yt^{3}/12\left(1-\nu^{2}\right)$ are the $2$D Young's modulus and $2$D bending modulus, respectively, with $t\approx 0.35\mathrm{nm}$ is the thickness of a mono-layer graphene. The Possion's ratio is reported around $\nu=0.16$~\cite{yue2012}. The integral is over the whole area of the spherical graphene bubble.

As an initial estimation, the average strain in a spherical bubble of graphene is about $\varepsilon_x=\varepsilon_y=1-\sin\theta_0/\theta_0$, and $\varepsilon_{xy}=0$ (the definition of $\theta_0$ is shown in
Fig.~\ref{fig:fig1},\cite{zong10} while the curvatures are $\kappa_x=\kappa_y=1/R$ and $\kappa_{xy}=0$, the elastic energy with this isotropic strain energy has a form of
\begin{equation}
\label{compareenergy}
E_{\mathrm{elas}}=D\frac{1+\nu}{R^2}\int\mathrm{d}^2s+
\frac{E_{\mathrm{2D}}}{1-\nu}\int\left(1-\frac{\sin\theta}{\theta}\right)^2\mathrm{d}^2s,
\end{equation}
thus the ratio of the two types of elastic energy is
$E_{\mathrm{curvature}}/E_{\mathrm{strain}}=\gamma\left(R/t\right)^2$, with $\gamma=12\left(1-\sin\theta_0/\theta_0
\right)^2$. Since $R\approx 30\mathrm{\mu m}$, $R/t\approx 10^5$, and $\theta_0\leq 0.3$,~\cite{georgiou11}
$E_{\mathrm{strain}}\gg E_{\mathrm{curvature}}$, only strain energy matters for graphene bubble observed in Ref.~\onlinecite{georgiou11}. The total strain energy is estimated to be approximately $E_{\mathrm{strain}}=\left(\pi/90\right)E_{\mathrm{2D}}R^2\sin^6\theta_0$, for $\theta_0\leq 0.3$.

A more accurate description of the elastic energy of the stretched carbon membrane with a bubble needs the knowledge of the strain distribution on the curved bubble surface. The axis-symmetric deflection profile of bubble was suggested to be approximately~\cite{yue2012}
\be\label{zr}
z(r)=h\left(1-r^2/\rho^2\right),
\ee
with $h$ the height of the spherical crown and $\rho$ the radius of the circular area of substrate without the coverage of graphene sheet. Equation~(\ref{zr}) is reasonable due to the fact that typically $\theta_0\ll 1$. The axis-symmetric distribution of radical displacement of the bubble is proposed as,
\be\label{ur}
u(r)=u_0\frac{r}{\rho}\left(1-\frac{r}{\rho}\right).
\ee

This strain distribution ensures that $u(\rho)=0$ and $u$ reaches its maximum at $r=0$. The elastic energy can be expressed by the radial and circumferential strains~\cite{yue2012},
\be\label{strainenergy}
E_{\mathrm{elas}}=2\pi\int_0^\rho r\mathrm{d}r \frac{E_{\mathrm{2D}}}{2\left(1-\nu^2\right)}\left(\varepsilon_r^2+
2\nu\varepsilon_r\varepsilon_{\theta}+\varepsilon^2_{\theta}\right).
\ee
Here, the radial strain is $\varepsilon_r=\left(u_0/\rho\right)
\left(1-2r/\rho\right)+2h^2r^2/\rho^4$, and the circumferential strain is
$\varepsilon_{\theta}=\left(u_0/\rho\right)\left(1-r/\rho\right)$.

Usually, separating a monolayer graphene adhering on the top of the silicon dioxide substrate costs adhesion energy per unit area $\gamma\approx 0.45\mathrm{J\cdot m}^{-2}$.~\cite{koenig11} However, in the Georgiou {\it et al.}'s experiment, the graphene layer can not stick to the rough substrate tightly, and there is also hydrophobic interaction caused by the present of water molecules in the gap between the graphene and rough substrate, which dramatically lowers the measured adhesion energy per unit area to $\Gamma_0\approx 0.1\mathrm{J\cdot m^{-2}}$.

The total potential energy for the graphene bubble has a form,
\be\label{totalpotential}
\Pi(\rho,h,u_0)=E_{\mathrm{strain}}-2\pi(p-p_0)\int_0^{\rho} z(r)r\mathrm{d}r.
\ee

With fixed radius $\rho$, the equilibrium condition $\partial\Pi/\partial u_0=\partial\Pi/\partial h=0$ leads
the potential energy~\cite{yue2012}
\be\label{reduce}
\Pi(\rho,N)=\frac{Nk_{\mathrm{B}}T}{4}-Nk_{\mathrm{B}}T\ln\frac{p_0\rho^{5/2}}
{(Nk_{\mathrm{B}}T)^{3/4}E_{2D}^{1/4}}.
\ee
Here, $N$ is the number of air molecules trapped in the bubble, which is constant due to the impermeability of graphene.\cite{bunch08} The first term in the right hand side of Eq.~(\ref{reduce}) is the strain energy stored in graphene, which is independent of the bubble geometry. The second term is the potential energy of the air molecules. As $\rho$ increases, the total potential energy decreases, while the adhesion energy increases due to part of the graphene is detached from the substrate. The equilibrium condition, 
\be\label{Gamma}
\left(\frac{\partial \Pi}{\partial \rho}\right)_N=-2\pi \rho \Gamma,
\ee
gives rise to the the adhesion energy per unit area~\cite{yue2012}
\be\label{Gamma2}
\Gamma=\frac{5Nk_{\mathrm{B}}T}{4\pi \rho^2}\propto\frac{E_{2D}h^4}{\rho^4}.
\ee

\begin{figure}[htb]
\includegraphics[scale=0.7]{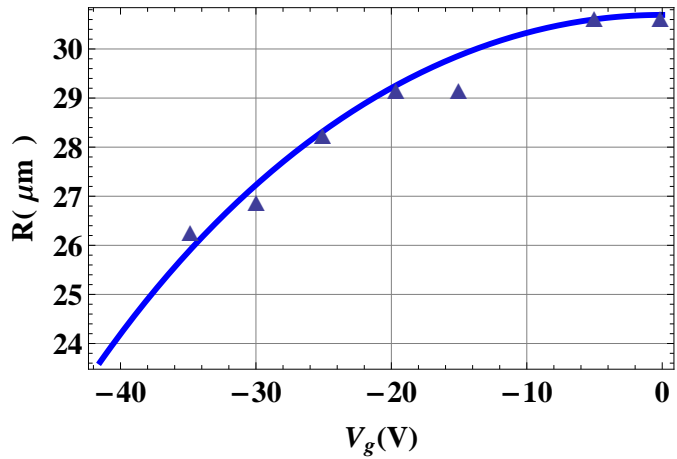}
\caption{\label{fig:fig2} The gate voltage dependence of radius of the graphene bubbles. The triangles denote the experimental data extracted from Ref.~[\onlinecite{georgiou11}]. The solid line is our theoretical relation, which is a part of an ellipse.}
\end{figure}

As we have emphasized, the MOS capacitance plays the key role in the electromechanical coupling of the graphene bubble. The simplest way to describe MOS capacitance is the parallel plate capacitor approximation, $c_g=\varepsilon_0\varepsilon/d_0$, with $d_0$ the thickness of the oxide layer. The gate voltage dependent radius of the graphene spherical bubble is obtained as,
\be\label{vgrrelation}
R=\sqrt{\frac{5Nk_{\mathrm{B}}T}{4\pi\sin^2\theta_0\Gamma_0}}\cdot\sqrt{1-\frac{\varepsilon_0\varepsilon V_g^2}{2\Gamma_0d_0}}.
\ee
Yue \emph{et al.}~\cite{yue2012} found that with the gate voltage changing from $-35\mathrm{V}$ to $0\mathrm{V}$, $Nk_{\mathrm{B}}T$ only varies a little around $170\mathrm{MeV}$ in Georgiou {\it et al.}'s experiment. The $\theta_0$, which is $1/2$ of the "contact angle" of the bubble, varies little too. In Table~\ref{tab1}, we list the $\theta_0$ for a set of negative gate voltages extracted from Ref.~\onlinecite{georgiou11}.
\begin{table}[htbp]
\caption{\label{tab1}
The $\sin^2\theta_0,(\theta_0\equiv \rho/R)$ for the gate voltage $-35$ to $0\, \mathrm{V}$.}
\begin{ruledtabular}
\begin{tabular}{lcccc}
$V_g$ (V) & -35 & -25 & -15 & 0\\
$ \sin^2\theta_0 $ & 0.072 & 0.068 & 0.066 & 0.065\\
\end{tabular}
\end{ruledtabular}
\end{table}
Equation (\ref{vgrrelation}) has a form $R=R_0 \sqrt{1-\alpha V_g^2}$. This relation are in good agreement with Georgiou \emph{et al.}'s experiment,~\cite{georgiou11} as shown in Fig.~\ref{fig:fig2}.
The fitted adhesion energy is $\Gamma_0=0.16\mathrm{J\cdot m^{-2}}$, which is in accordance with the result in Ref.~\onlinecite{yue2012}. The theoretical calculated thickness of the silicon dioxide is about $d_0\approx 400\mathrm{nm}$, which is also comparable to the experimental measurement of thickness.~\cite{georgiou_new} Therefore, our simple model captures the basic physics of the electromechanical coupling in the gate voltage controlled graphene bubble rising on the substrate.

In summary, we have theoretically investigated the gate voltage dependence of radius of the graphene spherical bubble observed by Georgiou {\it et al}. Base on the elastic membrane theory, we find this phenomenon is governed by the gate voltage dependent adhesion energy. In Georgiou \emph{et al.}'s experiment, the adhesion energy is sufficiently lowered by the water molecules trapped between the graphene and substrate. The key point of the gate voltage controlled expansion/ shrinkage of the graphene bubble is the adhesion energy modified by capacitive energy, which affects the mechanical balance of the equilibrium shape of the bubble. Present results can be used to find the design rules for the devices based on gate voltage controlled graphene bubble.

We gratefully acknowledge National Science Foundation of China (NSFC) under Grants No. 11074259, and the Major Research Plan of the National Natural Science Foundation of China (Grant No. 91027045) for the support of this work. We are grateful to Dr. Linying Cui for her modifying this manuscript.

\end{document}